# CHANCE AND NECESSITY IN FERMI'S DISCOVERY OF THE PROPERTIES OF THE SLOW NEUTRONS[1]

Alberto De Gregorio

*Doctorate of research in Physics at Università dell'Aquila*

*Museum and Department of Physics of Università di Roma "La Sapienza"*

In 1934 two important discoveries were made in Rome: at the beginning of March, while Frédéric Joliot and his wife Irène Curie had already observed in Paris the artificial radioactivity induced by α-particles, Enrico Fermi realised that also neutrons could be used to induce radioactivity; in fact, in spite of the weak intensity of the sources available at that time, electrical neutrality allowed these particles to approach the nucleus much closer than alphas did, and, if compared to the latter, passed through much thicker layers of matter.

Experiments carried out in Rome brilliantly confirmed the expectations, and Lord Ernest Rutherford personally complimented on them with these words:

*Dear Fermi, […] your results are of great interest […]. I congratulate you on your successful escape from the sphere of theoretical physics! You seem to have struck a good line to start with.*[2]

On October 22$^{nd}$ of that same year, Fermi discovered that, interposing some paraffin between the source of the neutrons and the sample, induced radioactivity increases sensibly.

The Indian astrophysicist Subrahmanyan Chandrasekhar reports that, at the beginning of the Fifties, Fermi himself described to him the circumstances of his own discovery:

---

[1] This paper derives from a study of the Proceedings of the seventh Solvay Conference, which I carried out while writing my thesis for the degree in Physics about "Fermi and the birth of theoretical physics of elementary particles: the β-decay theory". The thesis was supervised by professor Nicola Cabibbo and professor Fabio Sebastiani, and was discussed at the University "La Sapienza" of Rome on December 15$^{th}$ 2000.

[2] Rutherford's letter to Fermi of April 23$^{rd}$ 1934; quoted in E. Amaldi, *From the discovery of the neutron to the discovery of nuclear fission*, Physics Reports, **111,** nn**. 1-4** (1984), 1-332; here page 131.

*I described to Fermi Hadamard's thesis regarding the psychology of invention in mathematics, namely, how one must distinguish four different stages: a period of conscious effort, a period of "incubation" when various combinations are made in the subconscious mind, the moment of "revelation" when the "right combination" (made in the subconscious) emerges into the conscious, and finally the stage of further conscious effort. I then asked Fermi if the process of discovery in physics had any similarity. Fermi volunteered and said (his account made so great an impression on me that though this is written from memory, I believe that it is very nearly a truly verbatim account):*

*"I will tell you how I came to make the discovery which I suppose is the most important one I have made". And he continued: "We were working very hard on the neutron induced radioactivity and the results we were obtaining made no sense. One day, as I came to the laboratory, it occurred to me that I should examine the effect of placing a piece of lead before the incident neutrons. And instead of my usual custom, I took great pains to have the piece of lead precisely machined. I was clearly dissatisfied with something: I tried every "excuse" to postpone putting the piece of lead in its place. When finally, with some reluctance, I was going to put it in its place, I said to myself: 'No! I do not want this piece of lead here; what I want is a piece of paraffin'. It was just like that: with no advanced warning, no conscious, prior, reasoning. I immediately took some odd piece of paraffin I could put my hands on and placed it where the piece of lead was to have been."*[3]

Referring to the discovery that represents the first step towards the exploitation of the energy contained in the nucleus, Gerald Holton defines it a "mythological event"[4], all the more so because "everyone seems to have thought that the more energetic the neutrons, the greater would be their efectiveness".[5]

---

[3] S. Chandrasekhar, introduction to the article written with E. Fermi in 1953: *Magnetic field in spiral arms*. Contained in *E. Fermi. Note e Memorie (Collected Papers)*, 2 volumes, edited by Amaldi *et al.*, Accademia Nazionale dei Lincei and Chicago University, Rome-Chicago 1962-1965 (later on indicated as *F. N. M.*); here vol. II, pp. 926-927.

[4] G. Holton, *The scientific imagination. Case studies*, Cambridge University Press, Cambridge 1978; p. 174.

[5] Ibid., p. 173.

In the light of what has just been reported, the sudden action which allowed Fermi to discover the "miraculous effects of the filtration by paraffin",[6] as Emilio Segrè calls them, seems to be devoid of any rational and methodological justification and appears, instead, regulated by chance. In this paper I will try to reconstruct what – before October 22$^{nd}$ 1934 – was already known about the slowing down of neutrons by paraffin and about the dependence of their cross section on velocity, and how that knowledge could have an influence upon Fermi's discovery.

## § 1 – THE DISCOVERY OF THE EFFECTS OF HYDROGENATED SUBSTANCES ON THE RADIOACTIVITY INDUCED BY NEUTRONS

In April 1932, just two months after the observation of the neutron by Chadwick,[7] John Douglas Cockroft and Ernest Thomas Sinton Walton[8] were the first who saw the disintegrations provoked in atoms of light elements by artificially accelerated charged particles. The following month, the *Proceedings of the Royal Society* received a paper by Norman Feather, in which the English physicist demonstrated that nuclei disintegrate also if they are irradiated with neutrons.[9]

In his essay *From the discovery of the neutron to the discovery of nuclear fission*,[10] Edoardo Amaldi recalls that in January 1934 the Joliots announced the artificial radioactivity induced by α-particles, in the *Comptes Rendus de l'Académie des Sciences*; he also explains that

---

[6] E. Segrè, *Enrico Fermi. Physicist*, The University of Chicago Press, Chicago-London 1970; p. 80.

[7] J. Chadwick, *Possible Existence of a Neutron*, Nature, **129** (1932), 312.

[8] J. D. Cockroft, E. T. S. Walton, *Experiments with High Velocity Positive Ions II. Disintegration of Elements by High Velocity Protons*, Proc. Roy. Soc. A, **137** (1932), 229-242.

[9] N. Feather, *The Collisions of Neutrons with Nitrogen Nuclei*, Proc. Roy. Soc. A, **136** (1932), 709-727. The author had already observed the phenomenon the previous February and Rutherford himself had announced the result at the Royal Institution, on March 18$^{th}$; an abstract of Rutherford's report is in *Origin of the γ-Rays – Radiation from Beryllium and the Neutron*, Nature, **129** (1932), 457-458.

[10] E. Amaldi (1984), p.110.

*after the discovery of Joliot and Curie it was natural to think of producing new radioactive nuclides using bombarding particles different from alpha particles, as for example protons, deuterons and neutrons.*[11]

However natural thinking of neutrons could be, available intensities (about one hundred thousand times weaker than those of α-particles) made still doubtful the efficacy of those particles in inducing radioactivity. Neutrons, nevertheless, had manifest advantages over alphas, "all due to the electric neutrality"[12]

As early as 1920 when, in the *Bakerian Lecture*, Lord Rutherford claimed the existence of the neutron, he said that *its external field would be practically zero, except very close to the nucleus, and in consequence it should be able to move freely through matter.*[13]

Much thicker matter, therefore, can be activated by the neutrons than by the alphas; in fact, feeling the effects of the nuclear Coulomb field, α-particles are rapidly adsorbed whereas, for neutrons, the true experimental limitation is due to the number of electrons, emitted in β-decay, able to come out of the irradiated sample and reach the detector. If to all that we add the aptitude of neutrons to approach the nuclei and, consequently, interact with them, we can easily find out that the weak intensity of the sources is partly counterbalanced.

Starting from March 1934, researches on neutron-induced radioactivity were systematically carried out by Fermi and his collaborators, on atoms of all elements of increasing weights. The studies made by the Roman physicists are described in detail by Amaldi, who recalls that experiments were soon turned to better specify the vague results first obtained for neutron-induced radioactivity. Amaldi and Bruno Pontecorvo, however, as they came to this stage, had to face an incomprehensible variability of experimental results:

*It became apparent that the activation depended on the condition of irradiation. In particular* [... ,] *there were certain wooden tables* [...] *which had miraculous properties. As Pontecorvo noticed accidentally, silver irradiated on those tables*

---

[11] Ibid., p.115.

[12] Ibid., p.126.

[13] E. Rutherford, *Nuclear Constitution of Atoms*, Proc. Roy. Soc. A, **97** (1920), 374-400; here p. 396.

*gained more activity than when it was irradiated on the usual marble table. […] In order to clarify the situation, I started a systematic investigation.*[14]

It was then planned, in order to analyse the effect of the surroundings, to study the shielding of lead on the neutrons: they were partly deviated from their straight path from the source to the sample but partly, if emitted in different directions, turned again towards the sample.

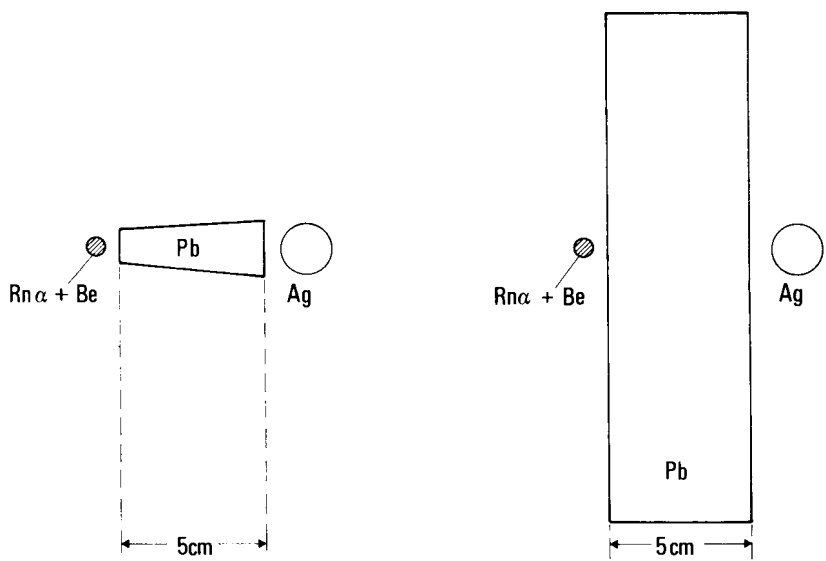

Fig. 1 – The shields of lead

The two configurations represented above in Amaldi's drawing were to clarify in which way some more lead (on the right) could compensate for the negative effect of a smaller shield (on the left).

Fermi hesitated for a long time in using some lead to filter the neutrons and, finally, he preferred to use paraffin, "with no conscious, prior, reasoning". In this way he discovered that the radioactivity induced by neutrons in the silver sample increases, by the very use of paraffin.

A similar account, supported by those by Fermi's collaborators, induced Holton to speak of a "mythological fact" when referring to this discovery. Amaldi, concerning it, recalls that

*the increase of the reaction cross section by reducing the velocity of the neutrons was at that time still contrary to our expectation*;[15]

---

[14] E. Amaldi (1984), p. 152.

while Segrè states that:

*We thought that the more energetic the neutrons the greater their effectiveness in producing reactions; how wrong we were we would discover only six month later.*[16]

That was a diffused prejudice, according again to Holton,[17] before the properties of slow neutrons were discovered in Rome; a prejudice perhaps justified by the actual weakening of the effects of charged particles when they slow down.

Amaldi, however, does not agree with the 'fatalist' version of discovery and states the importance of how systematically the experiments were conducted, up to that 22$^{nd}$ October.

It is decidedly right to acknowledge the Roman physicists' commitment and effort, but Fermi's intuition did not in fact depend on them.

The article that Fermi and collaborators wrote the day of the discovery thus starts:

*In the course of experiments on induced radioactivity in silver irradiated with neutrons, anomalies in the intensity of activation have been noticed; a piece of paraffin some centimetres thick, interposed between the source and silver, makes the activation increase instead of reducing it.*[18]

To what extent, nevertheless, can we really consider the observed result as completely unexpected? Could that "instead of reducing it" be referred to a comparison with lead? Better: what was already known - in 1934 - about the effects of paraffin on neutrons and about the cross section of the latter, and what did Fermi already know about all that, before discovering the radioactivity induced by slow neutrons?

---

[15] Ibid., p. 153.

[16] E. Segrè (1970), p.75.

[17] G. Holton (1978), p. 173.

[18] E. Fermi, E. Amaldi, B. Pontecorvo, F. Rasetti, E. Segrè, *Azione di sostanze idrogenate sulla radioattività provocata dai neutroni I*, Ric. Scient., **5, 2** (1934), 282-283; quoted in *F. N. M.*, 757-758.

## § 2 – THE RESEARCHES ABOUT NEUTRONS IN FRANCE AND IN THE REMAINDER OF EUROPE

The experiments, by which the penetrating radiation emitted by beryllium was established to be composed of neutrons, consisted in analysing the kinematics of the observed processes.

As the Joliots first recognised, that radiation is able to expel high energy protons from hydrogenous substances. In addition:

*Nous avons étudié ces rayonnements par l'ionisation qu'ils produisent dans une chambre montée sur un èlectromètre Hoffmann. […] Le courant augmente notablement quand on interpose des écrans de substances contenant de l'hydrogène comme la paraffine, l'eau, le cellophane. L'effet le plus intense a été observé avec la paraffine.*[19]

The two French physicists initially identified the penetrating radiation from beryllium with photons:

*Ces rayons γ de grande énergie sont capables de projeter des protons de grande vitesse quand ils traversent une substance hydrogénée. […] Cet effet s'accompagne d'une absorption considérable du rayonnement par les noyaux d'hydrogéne.*[20]

It was not possible, nevertheless, to refer to photons without self-contradictions. Chadwick carefully examined the kinematics of collisions in different elements and, in particular, the decrease of the energy transferred by the penetrating radiation of beryllium to atoms, as the mass of the latter increases; he established in this way that this radiation is made of neutral particles having about the same mass as protons.

Once it was understood that neutrons expel high energy protons from hydrogenous substances, this very property was used to recognise the presence of neutrons: "La chambre est tapisée intérieourement de paraffine de facon à détecter plus facilement les rayonnements de neutrons".[21]

---

[19] I. Curie, F. Joliot, *Émission de protons de grande vitesse par les substances hydrogénées sous l'influence des rayons γ trés pénétrants*, C.R. Acad. Sci., **194** (1932), 273-275; here p. 273.

[20] I. Curie, F. Joliot, *Effet d'absorption de rayons γ de très haute fréquence par projection de noyaux légers*, C.R. Acad. Sci., **194** (1932), 708-711; here p.709.

Maurice de Broglie and Louis Leprince-Ringuet studied the effects of screens, placed laterally with respect to the neutrons trajectory:

*Au cours des expériences on enregistre le nombre de passages de noyaux atomiques fortement ionisants, soit en mettant, soit en supprimant des écrans latéraux de plomb, paraffine, KCl. […] L'effet dû à la dispersion des neutrons par les écrans latéraux est considérable.*[22]

The effect of the lateral screens of lead is of the same order of magnitude as (for example) the walls of the laboratory but, "avec des écrans latéraux de KCl et de paraffine, il y a […] une forte augmentation du nombre des rayons observée": the two physicists, in fact, explain that the neutrons can expel atoms from the light elements composing the screens.

All these experiments can help us to understand the type of equipment used by Amaldi in Rome, named "castelletto"[little castle]:[23] it consisted of four little bricks of lead placed at the sides of a square, inside which the Italian physicist put a source of neutrons and a sample of silver to study the dependence of the activation on the mutual distance (this was the experiment which was carried out immediately before the idea of using the lead shields represented in figure 1); that activation changed only slightly, in contrast with what happened if the sample was outside the "castelletto".

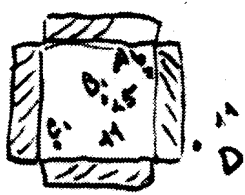

Fig. 2 – The "castelletto"

With regard to the properties of the neutrons, not only was their capability of expelling high energy protons known, but also that they are scarcely screened by lead.

---

[21] M. de Broglie, L. Leprince-Ringuet, *Sur la dispersion des neutrons du glucinium et l'existence de noyaux de recul provoqués par le lithium excité*, C.R. Acad. Sci., **194** (1932), 1616-1617; here p. 1617.

[22] Ibid.

[23] E. Amaldi (1984), p. 152.

In the session of the *Académie des Science* of May 9[th] 1932, for example, Jean Thibaud and F. Dupré la Tour reported that "au delà d'un écran de 30 cm de plomb, il subsiste encore plus du dixième de la radiation incidente".[24]

The two physicists analyzed the absorption of neutrons considering, in particular, the collisions against nuclei:

*Les choc nucléaires […] sont […] capables, s'ils ont lieu de plein fouet, d'absorber d'un coup la totalité de l'énergie du neutron (projection atomique) ou, au moins, de diffuser les neutrons et de les renvoyer vers l'arrière. Remarquons que le neutron, dépourvu de charge, peut, à la différence des corpuscules chargés, s'approcher très près de la surface nucléaire.*[25]

Thibaud and Dupré la Tour justified the high penetrating power of neutrons, due to their capability to approach practically undisturbed to the nuclei: "Pour un rayonnement préalablement fortement fitré, il n'y aurait absorption du neutron que dans les chocs extrêmement intimes avec les noyaux".[26]

The interest of the two physicists' report lies also in the mention of a different behaviour of slow neutrons with respect to fast neutrons, even though they explain this phenomenon in terms of a wider angular dispersion of slow neutrons:

*L'absorption comparativement plus intense qui a lieu dans les premiers centimètres de matière serait due à la filtration d'une partie beaucoup plus absorbable du rayonnement hétérogène (neutrons lents). On peut concevoir, par exemple, que ces neutrons lents subissent une diffusion latérale, sur les premières couches de matière, beaucoup plus intense que les corpuscules rapides.*[27]

The two French physicists confirmed the importance of the velocity of neutrons in one of their articles, the following October: "Nous pensons que les neutrons les moins pénétrants sont les plus fortement diffusés".[28]

---

[24] J. Thibaud, F. Dupré la Tour, *Sur le pouvoir de pénétration du rayonnement (neutrons) excité dans le glucinium par les rayons α*, C. R. Acad. Sci., **194** (1932), 1647-1649; here p. 1647.

[25] Ibid., pp. 1648-1849.

[26] Ibid., p.1649.

[27] Ibid.

[28] J. Thibaud, F. Dupré la Tour, *Sur l'affaiblissement de la radiation nucléaire du glucinium dans les écrans matériels*, C. R. Acad. Sci., **195** (1932), 655-657; here p. 656.

In the meanwhile, Feather obtained the disintegration of light elements resorting to neutrons. In spite of the weak intensity of incident radiation, the English physicist, using a cloud chamber, noticed a remarkable effectiveness in producing disintegrations:

*About 130 cases of interaction between a neutron and a nitrogen nucleus have been observed; of these about 30 resulted in disintegration, more than half of the latter without capture of the neutron. This is very different from the results obtained under α-particles bombardment, where elastic collisions […] outnumber inelastic (disintegrations) collisions by a factor of 1000:1. […] The former of these points of difference is certainly to be ascribed to the different extent of the external fields of the two particles.*[29]

The great efficacy of neutrons in inducing nuclear disintegrations was demonstrated by Feather in this way and, as we will soon see, would afterwards be also discussed by the Joliots. In his 1938 Nobel Lecture, while reconstructing the circumstances in which neutron induced radioactivity was discovered, Fermi himself would specify that "As a matter of fact, neutrons were already known to be an efficient agent for producing some nuclear disintegration".[30]

In July 1932, M. de Broglie and Leprince-Ringuet, in referring to the scarce absorption of neutrons by heavy elements, spoke of "trasparence du plomb",[31] which was then useful to filter neutrons and "diminuer les effets des rayons gamma de l'emanation". It was not lead, therefore, the most suitable element to shield the neutrons.

The Joliots dealt with important studies about neutrons. It is impossible to equivocate what is reported on page 23 of their paper *Preuves sperimentales de l'existence du neutron*:[32]

*L'étude de l'absorption du rayonnement Po + Be dans divers éléments montra une décroissance rapide de l'absorption massique en fonction du poids atomique de la*

---

[29] N. Feather (1932), p.723.

[30] E. Fermi, *Artificial radioactivity produced by neutron bombardment*, Nobel Lecture quoted in *F. N. M.*, vol. I, 1037-1043; here p. 1038.

[31] M. de Broglie, L. Leprince-Ringuet, *Sur les neutrons du bore excité par l'émanation du radium*, C. R. Acad. Sci., **195** (1932), 88-89; qui p. 89.

[32] I. Curie, F. Joliot, Journal de Physique, **4** (1933), 21-33; here p. 23.

*matière absorbante. Ce fait remarquable [...] permet de conclure que **tous les noyaux atomiques doivent être projetés, l'énergie qui leur est communiquée était d'autant plus grande que leur masse est plus faible, et que ce phénomène joue un rôle prépondérant dans l'absorption du rayonnement.***

Neither can be misunderstood what follows shortly after:

*La radiation de Po + Li est considérablement plus absorbée, à masse superficielle égale, par la paraffine par exemple, que par le plomb, contrairement à ce qui se passe pour les rayons γ du polonium;*[33]

precisely because of these foundations "il y a lieu de conclure à l'existence d'un rayonnement nouveau, probablement le neutron".

They soon turn to the absorption of neutrons in what follows:

*L'absorption d'un faisceau de neutrons se compose de deux termes. Un terme d'absorption vraie, correspondant à la perte d'énergie par choc sur un noyau (d'autant plus élevée que le noyau est plus léger), et d'un terme d'absorption par diffusion. [...] Ce dernier terme est prépondérant pour les noyaux lourds. [...] On observe globalement que les éléments légers produisent une absorption massique bien plus forte que les éléments lourds, caractère qui distingue le rayonnement de neutrons des autres radiations pénétrantes.*

As regards disintegrations by neutrons, the Joliots finally confirm what Feather stated in May:

*La probabilité de ce phénomène est grande relativement à celle des transmutations par les particules α. Ce fait tient à ce que le neutron peut facilement traverser la barrière de potentiel des noyaux.*[34]

This property partly counterbalances the weak intensity of neutrons flows, so much that we read in a note at the end of Joliots' paper: "D'après des expériences préliminaires effectuées à l'Institut du Radium de Vienne, de nombreuses substances émettraient des rayons α de transmutation sous l'action des neutrons".[35]

Summing up:

---

[33] Ibid., p. 25. The *masse superficielle* is the product of the density of the target multiplied by its thickness.

[34] Ibid., p. 32.

[35] Ibid.

*On voit donc que l'étude du phénomène d'émission des neutrons et celle des effets qu'ils produisent dans la matière peut être déjà considérée comme un des plus puissants moyens d'exploration des noyaux atomiques.*[36]

Another paper by the two French physicists can confirm the awareness of the efficacy that hydrogenous substances have in absorbing neutrons:

*On étudie le rayonnement nucléaire pénétrant excité par les rayons α du polonium dans les éléments légers en employant une chambre d'ionisation remplie de méthane ou de butane pour augmenter l'effet des neutrons; le rayonnement est extrêmement pénétrant et traverse 5 cm de plomb, sans absorption sensible. Par contre, l'absorption dans un bloc de 6,5 cm d'épaisseur de paraffine est notable.*[37]

When irradiating metallic sodium, "le rayonnement émis traverse 3 cm de plomb. Il est beaucoup plus absorbé dans 3 gr/cm$^2$ de paraffine que dans 3 gr/cm$^2$ de plomb, caractère qui indique la présence de neutrons".[38]

It is interesting, finally, if one recalls that Franco Rasetti was in Rome, what is then written:

*La courbe relative à l'excitation des neutrons du glucinium est en très bon accord avec celle de Rasetti qui a décelé les neutrons au moyen d'un compteur contenant de la paraffine.*[39]

There is no doubt, therefore, that in 1933 it was already known that light substances (like paraffin) slow down and absorb neutrons much more efficiently than heavy ones (like lead) do.

It was already known, moreover, that neutrons are a good means to induce nuclear disintegrations, due to their electrical neutrality, and that, as they assume different velocity, they show different behaviours.

The properties of the neutrons interacting with matter were studied more accurately - among others - by Chadwick and important conclusions about their scattering cross section were reached.

---

[36] Ibid.

[37] I. Curie, F. Joliot, *Nouvelles recherches sur l'émission des neutrons*, Journal de Physique, **4** (1933), 278-286; here pp. 278 e 279.

[38] Ibid., p. 279.

[39] Ibid., p. 282. Rasetti's quoted paper is *Über die Amergung von Neutronen in Beryllium*, Zeit. f. Phys., **78** (1932), 165-168.

## § 3 – THE STUDIES ABOUT THE NEUTRONS SCATTERING CROSS SECTION

On May 22$^{nd}$ 1933, more than one year before Fermi's experiments on slow neutrons, Chadwick gave the *Bakerian Lecture* entitled *The neutron*.[40] He discussed there some important characteristics of the particle he himself had discovered.

*The most obvious properties of the neutron are its ability to set in motion the atoms of matter through which it passes and its great penetrating power.*[41]

Chadwick, in addition, analyses "the dependence of the neutron emission on the velocity of the bombarding α-particles".[42] The neutrons are indirectly observed resorting to a ionisation chamber, with some paraffin at the entrance "to increase the effect".[43]

Chadwick thus concludes in reporting his results:

*The probability of a collision between a neutron and a nitrogen atom in the chamber or a proton in the paraffin wax depends on the velocity of the neutron.*[44]

The fifth paragraph closely examines the theory of *Collisions of neutrons with atomic nuclei*.[45] Chadwick specifies that the most interesting collisions are those between neutrons and protons and that, for these ones, experiments pointed out that "most of the collisions were due to slow neutrons".[46]

It is worth quoting the experiment to determine the "collision radius", that Chadwick now describes. The English physicist used a ionisation chamber filled with various gases (hydrogen, nitrogen, oxygen or argon) and he observed, in each case, the ionisation obtained by irradiating the gas with neutrons.

---

[40] J. Chadwick, *Bakerian Lecture – The Neutron*, Proc. Roy. Soc., **142** (1933), 1-25.

[41] Ibid., p.1.

[42] Ibid., p. 4.

[43] Ibid.

[44] Ibid., p. 5.

[45] Ibid., p. 15.

[46] Ibid., p. 17.

The experiment is carried out with neutrons both from beryllium and from boron. In the first case, the number of collisions in hydrogen is only a little lower than that in nitrogen and in oxygen; in the case of neutrons from boron (a bit slower),

*the number of deflexions in hydrogen was now about twice the number of deflexions in nitrogen, suggesting that the collision radius either of hydrogen or nitrogen varies rapidly with the velocity of the neutron. [...] Some experiments with slower neutrons suggest that the radius for the proton collisions continues to increase as the velocity of the neutron decreases.*[47]

Chadwick provides a theoretical explanation of the phenomenon, too: "In the consideration of the neutron-proton collision we thus have to explain [...] that the collision radius decreases".[48] The English physicist, to this end, quotes the expression of the cross section for two independent particles:

$$Q = \frac{h^2}{\pi M^2 v^2} \sum_n (2n+1) \cdot sin^2 \delta_n.$$

The zero order term is the only important one, which means that the range of the neutron-proton interaction is small compared with $\frac{h}{2\pi M v}$. As a consequence:

*The potential field of a proton and a neutron may be roughly likened to a very deep hole of small radius. [...] Thus $\delta_0 = \frac{1}{2}\pi$ approximately and $Q \approx \frac{h}{\pi M^2 v^2}$.*[49]

Chadwick thus obtains that the scattering cross section for neutron-proton collisions increases as the relative velocity *v* of the two particles decreases.

### § 4 – THE 1933 SOLVAY CONFERENCE

*In contrast to the conventional meetings in which reports are given on the successful solution of scientific problems, the Solvay Conferences were conceived to help directly in solving specific problems of unusual difficulty.*[50]

---

[47] Ibid., p.18.

[48] Ibid.

[49] Ibid., p.20.

In Werner Heisenberg's synthetic report, that is the style characterising the conferences promoted for the first time by the Belgian industrialist Ernest Solvay. The German physicist also adds that "there can be no doubt that [...] the Solvay Conferences played an essential role in the history of physics",[51] while referring particularly to the conferences before the war and, in a more specific way, to the one organised in 1933, which perhaps represented the last great opportunity for an international scientific debate before the war events.

## § 4.1 – THE POINT OF THE SITUATION

In the previous paragraphs we ascertained that, in 1933, hydrogenous substances were already known to better slow down the neutrons than lead does; in the same way as neutron-proton scattering cross section was acknowledged, both from experiments and from theory, as a decreasing function of the velocity of neutrons themselves.

It must be still clarified, however, what Fermi did know about the properties of paraffin and of slow neutrons.

The papers up to now quoted had all been published in prestigious reviews: *Journal de Physique*, *Proceedings of the Royal Society*, *Comptes Rendus de l'Académie des Sciences*, all having been signed by the most important nuclear physicists of the time. There is no reason to suspect that Fermi, leader of a group that had, for a long time, turned his own interest towards the atomic nucleus, was not up to date about those published results. Even if we admitted such a little reasonable possibility, more cogent arguments do exist, not only founded on good sense, that we can exactly connect with the Italian physicist's participation in the seventh Solvay Conference on *Structure et propriétés des noyaux atomiques*[52] which took place in Bruxelles from 22nd and 29th October 1934.

---

[50] W. Heisenberg, preface to J. Mehra *The Solvay Conferences on Physics. Aspects of the Development of Physics since 1911*, D. Reidel Publishing Company, Dordrecht – Holland/ Boston - U.S.A., 1975; p. V.

[51] Ibid., p. VI.

[52] *Structure et propriétés des noyaux atomiques*, Rapports et discussion du 7me Conseil de Physique tenu à Bruxelles du 22 au 29 octobre 1933, Gauthier-Villars, Paris 1934. The volume I used, borrowed from the library of the Physics department of the University "La Sapienza" of Rome, is the very one which belonged to Fermi.

The lectures where rich in new and stimulating contents; the Conference closely followed, in fact, some important experimental and theoretical achievements which provided plenty of topics for the physicists engaged in the study of nuclei and particles: the neutron, the positron and the deuteron joined to corpuscles already known, the existence of the neutrino was seriously considered and the nuclear forces represented an important way of description.

We are only going to analyse, among the remarkable amount of documents contained in the Proceedings of the Conference, the contributions of Chadwick and of the Joliots and the subsequent discussions: they witness the attention of the scientific community, and of Fermi himself, to what concerned neutrons.

### § 4.2 – CHADWICK'S CONTRIBUTION

In his speech, not only did Chadwick deal with *Diffusion anomale des particules α. Trasmutation des éléments par des particules α*, but he also treated of *Le neutron*,[53] devoting to that subject about one half of his talk. The English physicist largely reproposed what he had already presented in his *Bakerian Lecture*.

Chadwick first recalled that nuclear reactions take place only at distances comparable with nuclear dimensions, and then described the methods to determine the neutron mass, based on the knowledge of the kinetic energy of the neutron itself; the latter "peut être déterminée en misurant le parcours maximum des protons extraits de la paraffine".[54]

A long paragraph follows, in which *Collisions entre neutrons et noyaux atomiques*[55] are treated. Here are some passages:

*Le neutron ne sera dévié qu'en cas de chocs trés directs. […] La distribution des neutrons diffusés ne sera pas appréciablement dissymetrique. […] Il semble cepedant qu'en général des neutrons lents sont plus facilement diffusés que des neutrons rapides, et de Broglie a annoncé un effet analogue à l'effet Ramsauer pour les electrons.*[56]

---

[53] *Diffusion anomale des particules α. Trasmutation des éléments par des particules α. Le neutron*, ibid., 81-120.

[54] Ibid., p. 100.

[55] Ibid., p.103.

[56] Ibid., p. 104.

Chadwick quotes the theoretical studies carried out by Harrie Massey about collisions by neutrons and states that "les collisions entre neutron et proton présentent un intérêt tout particulier".[57] He also describes the results of experiments concerning neutron-proton scattering, specifying:

*La distribution angulaire des protons frappés par des neutrons a été observée par Auger et Monod-Herzen et aussi par Kurie. […] Les deux séries d'observations conduisent au même résultat: la distribution angulaire des protons par rapport au centre de masse du système en collision est à peu près uniforme. Le rayon de choc a été évalué par Meitner et Philipp. […] et il semble résulter de leurs données que la plupart des collisions étaient dues à des neutrons lents.*[58]

Then, Chadwick explains his own results:

*J'ai constaté que le rayon de choc de l'hydrogène varie avec la vitesse des neutrons. […] Quelques expériences avec des neutrons lents semblent indiquer que le rayon de choc continue d'augmenter quand la vitesse des neutrons diminue.*[59]

Chadwick, hence, clearly states that slow neutrons seem to have more chance to be scattered than fast ones. This is a statement stronger than that by Thibaud and Dupré la Tour, who only thought of a wider angular dispersion.

Chadwick now reports the formula for the scattering cross section, as obtained by the two independent particles collision theory of wave-mechanics:

$$Q \approx \frac{h^2}{\pi M^2 v^2}.$$

*La section de choc, pour une rencontre avec un proton, est grande et varie dans le sens voulu avec la vitesse du neutron.*[60]

### § 4.3 – THE JOLIOTS' CONTRIBUTION AND THE FOLLOWING DISCUSSION

The Joliots tackled the *Rayonnement pénétrant des atomes sous l'action des rayons α*.[61] Their talk mostly concerns the physics of the neutron, starting with the reconstruction of the discovery of that particle and going on with the study of its properties (among the others its absorption by matter) and of the characteristics of the γ-rays emitted, by some elements, together with neutrons. The second part is instead devoted to the study of *Électrons de matérialisation et de transmutation*.[62]

---

[57] Ibid., p. 105.

[58] Ibid., p. 105-106.

[59] Ibid., p. 106.

[60] Ibid., p.109.

[61] Ibid., pp. 121-156.

The debate following the Joliots' talk was very enlightening due to the quality and the number of those who partecipated in the discussion: Lise Meitner, Chadwick, Francis Perrin, Werner Heisenberg, Fermi, Maurice de Broglie, Walther Bothe, Ernest Lawrence, George Gamow, Rutherford, Rudolf Peierls, Niels Bohr, Wolfgang Pauli…

When Heisenberg takes the floor, he expresses doubts about the dependence of neutron-proton scattering cross section on the square of the de Broglie wavelength of neutrons,

*c'est-à-dire une aire qui surpasse la section du noyau; elle la surpasse de plus en plus à mesure que la vitesse des neutrons diminue. […] Il me semble que pour $r_0$ (rayon de la sphère d'action des noyaux) $<< \lambda$, la phase $\delta_0$ doit toujours être très petite.*[63]

Heisenberg will again tackle this argument but, before doing that, Fermi does talk about the cross section: he first "rappelle les hypothèses sur lesquelles est établie la formule donnée par M. Chadwick".[64] Therefore the Italian physicist, following Heisenberg, states that "les sections efficaces expérimentales sont plusieurs fois plus petites que celles donnée par la formule" and that "l'hypothèse d'une force d'échange ne suffit pas à améliorer la concordance".

Heisenberg than specifies "que la discordance entre la formule de Chadwick et les expériences n'est pas très grave".[65]

Maurice de Broglie, about some experiments concerning neutrons which seem not to agree with theory, moreover asks:

*Au sujet de l'interaction entre neutrons et atomes, que faut-il penser des courbes obtenues par Bonner, qui semblent prouver que l'absorption est fonction croissante de l'énergie des neutrons?*[66]

Chadwick answers that "dans le cas de l'hydrogène, la section est *grosso modo* inversement proportionnelle au carré de la vitesse du neutron".

---

[62] Ibid., pp.144-156.

[63] Ibid., p.161.

[64] Ibid.

[65] Ibid., p. 162.

[66] Ibid.

## § 5 – CONCLUSIONS

The experiments the Joliots carried out in France and those of Chadwick in Cambridge show the knowledge available, at the end of 1933, on the properties of neutrons passing through matter, and the Proceedings of the seventh Solvay Conference (which took place in October of that same year) show to what extent the international community was acquainted with those results. Fermi's argumentation in Bruxelles proves the interest of the Roman physicist for the behaviour of the neutrons. We can definitely state that, already at the end of 1933, he was aware both of the increase of the scattering cross section when the energy decreases and of the larger efficiency of paraffin with respect to lead in the slowing down and in the absorption of neutrons.

It is not difficult to suppose that in Fermi's subconscious mind some unconscious reasoning had started (since the time of his participation in the Solvay Conference, which began exactly one year before the experiments on slow neutrons, or since even previous circumstances) and that they resulted, after a "period of *incubation*" in the "immediate" decision he took in October 1934; that decision, in other words, would be the result of a subconscious elaboration of what was already known to the Italian physicist. Such a reconstruction would, among other things, confirm Hadamard's thesis on the psychology of inventions in mathematics and could be supported by the very fact that Fermi, in answering Chandrasekhar, thought of his 'sudden' decision to use paraffin as an example that could confirm the hypothesis of the great French mathematician.

In the light of what has here been exposed, it seems natural, if the lead wedge was used with the aim of studying the shield-effect on neutrons, that Fermi suddenly decided to substitute it with some piece of paraffin; and, in the same fashion, that he soon explained, after the lunch-break, the observed phenomenon as a consequence of the increase of the reaction cross section as the neutrons velocity decreases (even if the difference must be pointed out between the reaction cross section and the scattering cross section). Far from unexpected becomes, furthermore, what he wrote in his first paper together with the other Roman physicists: "It is plausible that the neutron-proton cross section grows up as the energy decreases";[67] while the sentence: "a piece of paraffin some centimetre thick, interposed between the source and the silver, makes the activation increase instead of reducing it"[68] could really imply a comparison with lead.

Fermi himself, anyway, while commenting with his collaborators upon the experiments which revealed the efficiency of the neutrons, slowed down by paraffin, in inducing radioactivity, said: "What a stupid thing to have discovered this phenomenon without having be able of foresee it".[69]

Giorgio Dragoni suggests, in an essay in which he looks for the warning signs of Fermi's discovery, that the reading of some of the Joliots' paper or, perhaps, the Italian physicist's participation in some seminars could really have some importance (but Fermi actually took part, much more than in a seminar, in the 1933 Solvay Conference). Dragoni himself specifies that

*putting in evidence this possibility does not mean to belittle the work of Fermi, of Amaldi and of the other members of the group at all. On the contrary, in this way we want to contribute to clarify and put in the right light what too often is labelled as a mere "chance" discovery.*[70]

---

[67] E. Fermi, E. Amaldi, B. Pontecorvo, F. Rasetti, E. Segrè (1934), p. 283 (*F. N. M.*, p.758).

[68] Ibid., p. 282 (*F. N. M.*, p.757).

[69] B. Pontecorvo, *Enrico Fermi*, Studio Tesi, Pordenone 1993; p. 82.

[70] G. Dragoni, *Un momento della vita scientifica italiana degli anni Trenta: la scoperta dei neutroni lenti e la loro introduzione nella sperimentazione fisica*, Physis, **18** (1976), 131-164; here p. 161. If someone would like to assign a role to chance in the discovery, he might refer to the fact that "Pontecorvo accidentally noticed" the dependence of the activation intensity on environmental conditions.